\documentstyle[a4,12pt]{article}
 
\begin{document}
\sloppy
 
\newcommand{\vv}{{\cal V}}
\newcommand{\dd}{{\rm deg}}
\newcommand{\de}{\delta}
\newcommand{\pr}{\prime}
 \newcommand{\0}{\bar{0}}
 \newcommand{\1}{\bar{1}}
 
\newcommand{\ds}{\displaystyle}
 
\newtheorem{theo}{Theorem}[section]
\newtheorem{lem}{Lemma}[section]
\newtheorem{prop}{Proposition}[section]
\newtheorem{cor}{Corollary}[section]


\hfill LPENSL-TH-15/98
\vskip 0.07truecm
\hfill MPI-PhT/98-94
\vskip 0.07truecm
\hfill LYCEN 98110
\vskip 0.10truecm
\hfill December 1998
 
\thispagestyle{empty}
 
\bigskip
\begin{center}
{\bf \Huge{Non-standard matrix formats}}
\end{center}
\begin{center}
{\bf \Huge{of}}
\end{center}
\begin{center}
{\bf \Huge{Lie superalgebras}}
\end{center}
\bigskip
\centerline{{\bf F. Delduc$^{\, a}$} $\ $ , $\ $
{\bf F. Gieres$^{\, b \, \S}$} $\ $,$\ $ 
{\bf S. Gourmelen$^{\, c}$} $\ $$\ $ 
and $\ $
{\bf S. Theisen$^{\, d}$}}
\bigskip
\centerline{$^a${\it Laboratoire de Physique}$^{\,
\dagger }$} 
\centerline{\it Groupe de Physique Th\'eorique ENS Lyon} 
\centerline{\it 46, all\'ee d'Italie}
\centerline{\it F - 69364 - Lyon CEDEX 07}
\bigskip
\centerline{$^{b}${\it Max-Planck-Institut f\"ur Physik}}
\centerline{\it - Werner Heisenberg Institut -}
\centerline{\it F\"ohringer Ring 6}
\centerline{\it D - 80805 - M\"unchen}
\bigskip
\centerline{$^{c}${\it Institut de Physique Nucl\'eaire}}
\centerline{\it Universit\'e Claude Bernard (Lyon 1)}
\centerline{\it 43, boulevard du 11 novembre 1918}
\centerline{\it F - 69622 - Villeurbanne CEDEX}
\bigskip
\centerline{$^{d}${\it Sektion Physik der Universit\"at M\"unchen}}
\centerline{\it Theresienstra\ss e 37}
\centerline{\it D - 80333 - M\"unchen}
\bigskip
\bigskip
 
\begin{abstract}
 
The standard format of matrices 
belonging to Lie superalgebras 
consists of partitioning the matrices into even and
odd blocks. 
In this paper, we study other possible matrix formats 
and in particular the so-called diagonal format 
which naturally occurs
in various applications, e.g. in superconformal field theory, 
superintegrable models, for super $W$-algebras 
and quantum supergroups. 
\end{abstract}

\nopagebreak
\begin{flushleft}
\rule{2in}{0.03cm} \\

{\footnotesize \ ${}^{\dagger}$
UMR 8514, CNRS et \'Ecole Normale Sup\'erieure de Lyon.}
\\  [-0.04cm]
{\footnotesize \ ${}^{\S}$
On leave of absence from Universit\'e Claude Bernard$^{\, c}$.}

\end{flushleft}

\newpage

\thispagestyle{empty}
\tableofcontents

\newpage
\setcounter{page}{1}

\section{Introduction}

Lie superalgebras of matrices are usually introduced by considering a
matrix representation of endomorphisms of a graded vector space. 
The standard matrix format 
 consists of arranging the even and odd
matrix elements
(i.e. matrix elements associated to even and odd
endomorphisms) into block form \cite{vk}-\cite{fss}.
In references \cite{gt,dm,gg},
we encountered a matrix format of the Lie superalgebras
$sl(n+1|n)$ and $osp(2m\pm 1|2m)$ 
in which there are even and odd diagonals, i.e.
alternatively even and odd elements in each row and 
column. (Particular 
examples of this arrangement 
also appeared in previous work concerning conformal 
field theory \cite{bsa,wl} or quantum groups \cite{qg}
and the infinite dimensional case was considered
for studying Fock space representations in quantum field theory
\cite{kac}.)
In references \cite{gt} and \cite{gg}, 
this format was
referred to as diagonal grading `representation' or
`realization', but to be more precise we will rather refer to
it as {\em diagonal matrix format} (as opposed to the {\em
block} or {\em standard matrix format} \cite{man}). 
For instance, a generic element of $M \in osp(3|2)$
can be parametrized in the following way in block and diagonal 
format, respectively : 
\begin{equation}
\label{1}
M_{{\rm block}} = \left[
\begin{array}{rrrrr}
{\bf -A} & {\bf C} & {\bf 0} & -\tau & -\mu \\
{\bf B} & {\bf 0} & {\bf -C} & \lambda & \beta \\
{\bf 0} & {\bf -B} & {\bf A} & -\alpha & \varepsilon \\
\varepsilon & \beta & -\mu & {\bf i} & {\bf -j} \\
\alpha & -\lambda & \tau & {\bf -k} & {\bf -i}
\end{array}
\right] 
\ \ , \ \ 
M_{{\rm diag.}} = \left[
\begin{array}{rrrrr}
{\bf A} & \alpha & {\bf B} & \varepsilon & {\bf 0} \\
\mu     & {\bf i} & \beta & {\bf j} & \varepsilon \\
{\bf C} & \lambda & {\bf 0} & -\beta & {\bf B}  \\
\tau     & {\bf k} & \lambda & {\bf -i} & -\alpha \\
{\bf 0} & -\tau & {\bf C} & \mu & {\bf -A}
\end{array}
\right] 
\ \ .
\end{equation}
Here, the boldface
entries (i.e. the Latin characters and zeros)
represent the even part of the matrix which part belongs to the 
ordinary Lie algebra 
$so(3) \oplus sp(2)$; the $so(3)$-submatrix of 
$M_{{\rm block}}$ 
(parametrized by capital Latin characters and zeros) 
is antisymmetric with respect to its antidiagonal 
for our choice of the metric (cf. eq.(\ref{supm})). 

In order to characterize the matrix $M_{{\rm diag.}}$ of eq.(\ref{1})
and more 
generally the elements $M \in osp(2m\pm 1 | 2m)$ 
in diagonal 
format, one labels the matrix diagonals by integer numbers, 
the main diagonal being counted as the $0$-th one. 
Then, the even part of $M$ only has entries 
on the even diagonals and these diagonals are alternatingly antisymmetric 
and symmetric with respect to the antidiagonal; 
the same applies to the odd diagonals \cite{gt}. 
Thus, the diagonal format leads to 
very simple and transparent expressions which are to be derived 
and discussed in the present paper.

Our discussion is based on references \cite{gt,dm, gg}
which were devoted to various applications of Lie superalgebras
to physics. 
In fact, the diagonal matrix 
format appears in a natural way in physics 
in the context of two-dimensional superconformal theories and
superintegrable models and in particular in the study of 
$W$-superalgebras. 
The latter and the related super Toda theories are defined 
in terms of basic Lie superalgebras
which admit an $osp(1|2)$ principal embedding\footnote{A classical
Lie superalgebra is called basic if it admits a non-degenerate 
invariant bilinear form \cite{fss}. While the study 
of the structure of ordinary Lie algebras and their classification
can be done by investigating the embeddings 
of $sl(2)$ into these algebras, the one of 
Lie superalgebras can be performed by considering the embeddings of 
$osp(1|2)$.}
\cite{lss,drs}: they are given by 
($n\geq 1$) 
\begin{eqnarray}
&&
sl(n+1|n) 
\quad , \quad 
sl(n|n+1) 
\nonumber 
\\
&&
osp(2n\pm 1|2n) \quad , \quad
osp(2n|2n) \quad , \quad
osp(2n+2|2n)  
\label{class}
\\
&&
D(2,1; \alpha ) \qquad {\rm with} 
\ \ \alpha \neq 0,-1 
\ \ .
\nonumber 
\end{eqnarray}
These algebras all admit a completely fermionic
system of simple roots and are therefore referred to as 
{\em fermionic Lie superalgebras} \cite{drs}. 
If the diagonal format is adopted, 
the Chevalley generators 
associated to these simple fermionic roots 
can be chosen to be matrices 
which only have non-vanishing 
entries on the first upper diagonal;   
e.g. in expression (\ref{1}), the two Chevalley generators 
are parametrized by  
$\alpha$ and $\beta$, respectively.   
This choice of generators is quite analogous to the one that is generally 
considered for ordinary Lie algebras like $sl(n)$ 
because of its simplicity 
and convenience.  
We also note that the expressions for the supertrace, grading,... 
are easier to manipulate for the diagonal format than the standard one 
since they do not 
refer to a partitioning of matrices by blocks. 

Our paper is organized as follows.  
After recalling some 
general notions concerning Lie superalgebras in section 2, we study 
their possible matrix formats   
for which we determine 
the corresponding expressions for the
graded trace, commutator or transpose (sections 3 and 4).
Thereafter, we focus on the diagonal format and 
discuss explicit results for the Lie superalgebras   
listed in equation (\ref{class}) as well as for their infinite dimensional 
limit $(n \to \infty)$. 
We conclude with some comments on the graded structure of the root space 
of $sl(n|m)$.

\section{Lie superalgebras}

In this section, we summarize some basic facts about 
Lie superalgebras \cite{vk}-\cite{fss}
and we recall some definitions concerning fermionic 
Lie superalgebras \cite{drs, fss}.
No reference will be made to   
matrix representations which will only be discussed
in the subsequent sections. 

\subsection{Generalities}

We consider finite-dimensional Lie superalgebras over the
field ${\bf C}$  of complex numbers. 
(The generalization from complex numbers to Grassmann numbers
considered in some of the references \cite{vk}-\cite{fss}
is fairly straightforward and will not be discussed here.) 
{\em Graded} will always mean
${\bf Z}_2$-graded and the elements of ${\bf Z}_2$ will be 
denoted by $\0$ and $\1$.

Consider a graded, complex  vector space
$V = V_{\0} \oplus V_{\1}$
with ${\rm dim} \, V_{\0}=n,  \, {\rm dim} \, V_{\1}=m$ and
$n\geq 1, m \geq 1$ finite. The elements of $V_{\0}$ and
$V_{\1}$ (i.e. the {\em homogeneous} elements of $V$) 
are referred to as {\em even}
and {\em odd}, respectively.  This 
grading of $V$ 
amounts to the definition of an {\em involution} operator 
on $V$ (which is also referred to as the 
{\em parity automorphism}):  this 
map is defined for  
the homogeneous elements of $V$ by 
\begin{eqnarray}
\label{invo}
\epsilon \ : \ V 
& \! \! \! \! & 
\to \ \ V
\\
v 
& \! \! \! \! & 
\mapsto \ \ \epsilon (v) = (-1)^{\dd \, v} v
\qquad {\rm with} \quad
\dd \, v = \left\{
\begin{array}{c}
0 \quad {\rm if} \ \, v\in V_{\0} \\
1 \quad {\rm if} \ \, v\in V_{\1}
\end{array}
\right.
\nonumber
\end{eqnarray}
and extended to the whole of $V$ by linearity. 
It satisfies $\epsilon^2 ={\rm id}_{V}$.

A {\em graded endomorphism} $M$ 
of $V$ is an endomorphism which is compatible 
with the grading of $V$, i.e. for every $k\in \{ \0 , \1 \}$, we have
$$
M V_{k} \subset V_{k} 
\qquad {\rm or}  \qquad 
M V_{k} \subset V_{k +\1}
 \ \ . 
$$
To these endomorphisms one 
assigns parities ${\rm deg} \, M=0$ and ${\rm
deg} \, M=1$, respectively, and refers to them as
{\em even} and  {\em odd}.  
The composition of any two endomorphisms $M$
and $N$ is  denoted by $MN$.

The grading of $V$ induces a similar grading of 
the vector space ${\rm End} \, V$ of all endomorphisms of
$V$: ${\rm End} \, V = {\rm End}_{\0}  \, V  \oplus {\rm End
}_{\1} \, V$
where ${\rm End}_{\0} \, V$ and ${\rm End}_{\1} \, V$ 
denote the
 vector spaces of even and odd endomorphisms, respectively
(i.e. the graded endomorphisms are the homogeneous 
elements of ${\rm End}\, V$). 
The grading corresponds to  
an involution operator which is defined by 
\begin{eqnarray}
\label{in}
{\rm Ad}_{\epsilon} \ : \  {\rm End} \, V 
& \! \! \! \! &
\longrightarrow  \ \ {\rm End} \, V
\\
M \ \ 
&\! \! \! \!  & 
\longmapsto \ \
{\rm Ad}_{\epsilon} \, M = \epsilon M \epsilon^{-1}
\ \ . 
\nonumber
\end{eqnarray}
It changes the signs of all odd elements of ${\rm End}
\, V$ and leaves the even elements invariant, henceforth
$({\rm Ad} _{\epsilon})^2 = {\rm id} _{{\rm End}
\,V}$. 

The {\em graded commutator} is introduced 
for graded endomorphisms $M,\, N$ by 
\begin{equation}
\label{gc}
[M, N \} = M  N -
(-1)^{(\dd \, M)(\dd \, N)} N  M
\end{equation}
and extended to the other endomorphisms by
bilinearity. 

The {\em supertrace} of $M \in {\rm End} \, V$ is defined by
\begin{eqnarray}
{\rm str} \ : \ {\rm End} \, V 
& \! \! \! \! & 
\to \ \ {\bf C} 
\nonumber \\
M 
& \! \! \! \! & 
\mapsto \ \ {\rm str} \, M = {\rm tr} \, \epsilon M
\ \ ,
\label{tr}
\end{eqnarray}
where the ordinary trace `tr' is defined by means of an
arbitrary matrix realization. (For a coordinate-free
definition, we refer  to \cite{man, cons}.)
The homomorphism (\ref{tr}) has the fundamental property
\begin{equation}
{\rm str} \, [M, N \} =0
\ \ .
\end{equation}
 
By definition, 
the general linear Lie superalgebra
$gl(n| m)$ is the graded vector space ${\rm End} \, V$ 
together with the graded commutator (\ref{gc}). 
The Lie superalgebra
$sl(n | m)$
consists of all elements of $gl(n | m)$ 
with vanishing supertrace.

The definition of the subalgebras of $sl(n | m)$
refers to the notion of supertransposition.
The {\em supertranspose} $M^{\ast}$ of a graded 
endomorphism  
$M \in {\rm End} \, V$ is defined
by its action on homogeneous elements
$\omega \in V^{\ast}$ ($V^{\ast}$ being 
the graded vector space which is dual to $V$):
\begin{equation}
\label{cofree}
\left( M^{\ast} (\omega) \right) (v) =  (-1)^{({\rm deg} \, M)({\rm deg} \,
\omega)} \  \omega \left( M (v) \right)
\qquad \mbox{for all} \ \ v \in V 
\ \ . 
\end{equation}

\subsection{On the structure of fermionic Lie superalgebras}

\subsubsection{Chevalley basis}

For any basic Lie superalgebra 
$\cal G$, there exists a unique (up to a 
constant factor) bilinear 
form $\langle \ , \ \rangle$ on $\cal G$ that 
has the following properties \cite{vk, fss}: 
it is consistent\footnote{A bilinear form 
$\langle \ , \ \rangle$ on ${\cal G} = {\cal G}_{\0} \oplus {\cal G}_{\1}$
is called consistent if $\langle X, Y \rangle =0$ for all 
$X \in {\cal G}_{\0}$ and $Y \in {\cal G}_{\1}$.}, non-degenerate, 
graded symmetric and invariant. 
For all fermionic Lie superalgebras except 
$osp(2n+2|2n)$ and $D(2,1; \alpha )$, this form 
coincides with a multiple of the 
Killing form, i.e. 
$\langle M , N \rangle = {\rm str} ({\rm ad}_M  \, {\rm ad}_N) $. 

Consider a Cartan subalgebra $\cal H$ of $\cal G$ 
and a system of simple roots with respect to $\cal H$.
For fermionic Lie superalgebras to which we restrict 
our attention here, the simple root system 
can be chosen as completely fermionic,
i.e. the endomorphisms associated to these roots can all be chosen 
as odd.
We will come back to other possible choices in the concluding section.

For $i=1,\dots, {\rm rank}\, {\cal G}$, let $h_i$
parametrize a basis of the Cartan subalgebra and 
let $e_i, f_i$ denote odd endomorphisms
associated to simple roots and their negatives, respectively. 
A {\em Chevalley basis} of ${\cal G}$ consists of a set of
generators $h_i, e_i, f_i$ 
which satisfy the commutation relations  
\begin{eqnarray}
{[h_i , h_j ]} &=& 0 
\qquad \ \; , \quad 
{[h_i , e_j ]} \ =\ +  a_{ij}e_j
\nonumber 
\\
\label{algebra}
\{ e_i,f_j \} &=& \delta_{ij} h_j
\quad , \quad 
{[h_i , f_j]} \ =\ -a_{ij} f_j
\ \ ,  
\end{eqnarray}
where $a_{ij} \equiv \langle h_i , h_j \rangle  \, / \,  
\langle e_j , f_j \rangle$ defines the elements of the 
{\em Cartan matrix} and where $\langle e_i , f_j \rangle =0$ 
for $i \neq j$.
The endomorphisms $h_i$ and $e_i, f_i$ are referred 
to as {\em Cartan} and {\em Chevalley generators},
respectively. 

By the rescaling 
\[    
e_i ^{\prime} =  \alpha_i e_i \quad , \quad  
f_i ^{\prime} = \beta_i f_i \quad , \quad 
h_i ^{\prime} =  \alpha_i \beta_i   h_i 
\qquad {\rm with} \ \; 
\alpha_i \beta_i \equiv 
 \langle e_i , f_i \rangle    ^{-1} 
\ \ ,
\]
one can always achieve that the Cartan matrix 
is symmetric and given by $a_{ij} ^{\prime} =  
\langle h_i^{\prime} , h_j^{\prime}
\rangle$. We note that there are other transformations 
which preserve the defining relations (\ref{algebra}) of the Chevalley
basis, e.g. the rescaling $(h_i, e_i, f_i) \to 
(h_i, q e_i, {1 \over q} f_i)$ with $q \in {\bf C}^{\ast}$ 
or the 
graded extension of the Chevalley involution \cite{rs, vk}:
\[
\sigma (h_i ) = - h_i 
\quad , \quad  
\sigma (e_i ) = - f_i 
\quad , \quad  
\sigma (f_i ) =  e_i
\qquad (\; \sigma^4 = {\rm id} \; ) 
\ \ . 
\]

\subsubsection{Principal $osp(1|2)$-embedding}

An embedding of the Lie superalgebra 
$osp(1|2)$ into $sl(n|n+1)$ is a non-trivial 
homomorphism from $osp(1|2)$ into $sl(n|n+1)$.
The {\em principal embedding} which is denoted by 
$osp(1|2)_{\rm pal}\subset sl(n+1|n)$ and defined 
up to conjugation, is distinguished by the fact that it leads
to the so-called principal gradation \cite{rs}; 
it is explicitly given by \cite{drs, fss,lss} 
\begin{equation}
\label{spal}
J_- := \sum_i f_i 
\qquad , \qquad 
J_+  :=  \sum_{i,j}a^{ij}e_i
\qquad , \qquad 
H := \{ J_+,J_- \}
=  \sum_{i,j}a^{ij} h_i
\ \ ,  
\end{equation}
where the $a^{ij}$ denote the elements of the inverse
Cartan matrix.   
Together with the  bosonic generators
$X_{\pm} :=  {1 \over 2} \{ J_{\pm}, J_{\pm} \}$, 
the endomorphisms (\ref{spal}) define a basis of 
the Lie superalgebra $osp(1|2)$: 
in fact, the 
commutation relations 
${[H,e_i ]} =e_i $ and 
${[H,f_i ]} =-f_i$ 
imply 
\begin{eqnarray}
{[H, J_{\pm} ]}  & = & \pm J_{\pm}
\qquad \!  , \quad
\{ J_+ , J_- \} \, = \, H 
\\
{[ H, X_{\pm} ]} &=& \pm 2X_{\pm}
\quad , \quad
{[ X_+, X_- ]} = - H
\quad , \quad
{ \lbrack J_\mp, X_\pm\rbrack } =\pm J_\pm
\ \ .
\nonumber 
\end{eqnarray}

\section{$gl(n| m)$ in standard matrix format}
 
\subsection{Definition}
    
Consider a basis $\{e_1,..., e_n \}$ of $V_{\0}$
and a basis $\{e_{n+1},..., e_{n+m} \}$ of $V_{\1}$ 
\footnote{The basis vectors $e_1,...,e_{n+m}$ 
of the vector space $V$ 
should not be mixed up with 
the Chevalley generators (introduced in the previous section
and further discussed in section 6)
which are endomorphisms of $V$.}.
Choose an ordering $(e_1, ..., e_{n})$ of the first basis 
and an ordering $(e_{n+1}, ..., e_{n+m})$ of the second one. 
Any vector $v \in V$ can be expanded with respect to the 
{\em ordered homogeneous basis} $(e_1, .., e_n, e_{n+1},..,
e_{n+m})$ of $V$:
\[
v = v_1 e_1 + ... + v_{n+m}
e_{n+m}
\ \ .
\]
The associated column of complex numbers 
will be denoted by the same
 symbol: $v = \left[ v_1, ..., v_{n+m} \right]^T$. 
(Here, the superscript
`$T$' denotes the ordinary transpose.)

Similarly, an endomorphism $M$ of $V$ can be expanded 
with respect to the 
ordered homogeneous basis $(e_1, ...,e_{n+m})$:
\[
M(e_j) \equiv 
\sum_{i=1}^{n+m} M_{ij} e_i \equiv 
\left\{ 
\begin{array}{lll}
\sum_{i=1}^{n} A_{ij} e_i +
\sum_{i=n+1}^{n+m} B_{ij} e_i
&\quad & {\rm for} \ j \leq n 
\\
& & 
\\
\sum_{i=1}^{n} C_{ij} e_i +
\sum_{i=n+1}^{n+m} D_{ij} e_i
&\quad & {\rm for} \ j > n \ \ . 
\end{array}
\right.
\]
Thus, with respect to the given basis, the endomorphism
$M$ corresponds to  
an $(n+m) \times (n+m)$ matrix with
complex components which we also denote by $M$: 
\begin{equation}
\label{bl}
M = 
\left[
\begin{array}{cc}
A   & B \\
C   & D
\end{array}
\right]
= 
\left[
\begin{array}{cc}
A   & 0 \\
   0& D
\end{array}
\right]
+
\left[
\begin{array}{cc}
0   & B \\
C   & 0
\end{array}
\right]
\ \ .
\end{equation}
Here, the matrix involving $A_{n\times n}$ and $D_{m\times m}$
defines the even part of $M$ and the one involving
$B_{n\times m}$ and $C_{m\times n}$ its odd part. 

Thus, the grading of vectors $v \in V$
and the induced grading of endomorphisms $M \in {\rm End} \, V$ 
can be implemented by  even and odd blocks
for the representative column vectors and matrices. 
This format of column vectors and matrices is called the {\em
block} or {\em standard
format} \cite{man}; 
it is defined up to even changes of basis 
- see section 3.3 below. 

From now on, $v$ and $M$ will always denote the representative
column vectors and matrices, respectively, 
and not the abstract vectors and endomorphisms
discussed in the previous section.

\subsection{Some explicit expressions}

For convenience, one often says that the component
$v_i$ of a vector
is even or odd if the vector all of whose entries  but $v_i$ 
are zero, is even or odd. 
(Of course, in the present context,
this is an abuse of terminology
since $v_i$ simply represents a complex number.
However, $v_i$ has a Grassmann parity 
if one considers
the tensor product of the complex vector space $V$ with a Grassmann
algebra as often done in physics.) 
Similarly the element
$M_{ij}$ of a matrix $M$ is said to be even or odd if 
it belongs to the even or odd part of the matrix.
In this sense, 
the grading of the vector and matrix components is given 
in terms of 
the {\em parity} $\alpha(i)$ {\em of the label} $i$,  
\[
\alpha (i) := {\rm deg} \,  e_i
\qquad {\rm for} \quad i \in \{ 1,...,n+m \}
\ \ ,
\]
by 
\begin{eqnarray} 
{\rm deg} \,  v_i & = & \alpha (i) 
\nonumber 
\\
{\rm deg} \, M_{ij} & = & \alpha (i) + \alpha (j) 
\ \; {\rm mod} \, 2 
\ \ .  
\end{eqnarray}

For the standard matrix format, the parity automorphism
$\epsilon$ is represented by 
\begin{equation}
\label{ep}
\epsilon = \left[
\begin{array}{cc}
{\bf 1}_{n}  & 0 \\
0 & -{\bf 1}_m
\end{array}
\right]
\ \ ,
\end{equation}
the supertrace takes the explicit form
\begin{equation}
\label{bstr}
{\rm str} \, \left[
\begin{array}{cc}
A & B \\
C & D
\end{array}
\right]
= {\rm tr} \, A - {\rm tr} \, D
\end{equation}
or
\begin{equation}
{\rm str} \, M  = \sum_{i=1}^{n} M_{ii} -
\sum_{i=n+1}^{n+m} M_{ii} 
= \sum_{i=1}^{n+m} (-1)^{\alpha (i)} \, M_{ii}
\ \ .
\end{equation}
 
The automorphism
${\rm Ad}_{\epsilon}$
of  ${\rm End} \, V$ reads 
\begin{equation}
{\rm Ad}_{\epsilon} \, M = \epsilon M \epsilon^{-1}=
\left[
\begin{array}{cc}
A & -B \\
-C & D
\end{array}
\right]
\ \ ,
\end{equation}
and thus amounts to changing the signs of all odd entries
of $M$.

From the coordinate-free definition (\ref{cofree})
of the supertranspose $M^{\ast}$ 
of an endomorphism $M$, one can 
easily deduce a matrix expression \cite{scheu}. 
However, the usual 
definition of the {\em supertranspose} $M^{sT}$ of a matrix $M$ 
differs from this expression  
by the automorphism ${\rm Ad}_{\epsilon}$, i.e.
$M^{sT} = {\rm Ad}_{\epsilon} \, M^{\ast}$; more explicitly,
\begin{equation}
\label{sst}
M^{sT} = \left[
\begin{array}{cc}
A^T & C^T \\
-B^T & D^T
\end{array}
\right]
\end{equation}
or 
\begin{equation}
\left( M^{sT} \right) _{ij} = M_{ji} \, 
(-1) ^{\alpha(i) (\alpha (j)
+1)} \ \ .
\end{equation}
This operation satisfies
\begin{equation}
\label{2}
\left( M^{sT} \right) ^{sT} = {\rm Ad}_{\epsilon} \, M
\end{equation}
and, for the homogeneous elements, 
\begin{equation}
\label{mn}
\left( M  N \right)  ^{sT} =
(-1)^{(\dd \, M)(\dd \, N)} N^{sT} M^{sT}
\ \ ,
\end{equation}
which relation implies
\begin{equation}
\label{cm}
\left[ M , N \right\} ^{sT} =
- [ M^{sT}, N^{sT} \}
\ \ .
\end{equation}

In summary, the standard or block format  of
$gl(n | m)$
consists of matrices (\ref{bl}), 
the grading, supercommutator, supertrace and supertranspose being
defined by eqs.(\ref{ep}),(\ref{gc}), (\ref{bstr}) and (\ref{sst}),   
respectively.

\subsection{Formats and changes of formats} 

In order to define matrix formats in general terms, we   
consider once more 
the standard format for column vectors and matrices 
discussed in the previous section.   
Under a change of basis, the vector $v$ transforms according 
to $v \mapsto v^{\prime} = F v$ where $F$ denotes a non-singular 
matrix (i.e. the superdeterminant of $F$ does not vanish).    
The induced change of matrices is a similarity transformation, 
$M \mapsto M^{\prime} = FMF^{-1}$, in particular 
$\epsilon \mapsto \epsilon ^{\prime} = F\epsilon F^{-1}$. 

Obviously, 
the only changes of basis which 
do not modify the partitioning of matrices 
into even and odd blocks are  
separate changes of basis in $V_{\0}$ and  $V_{\1}$.
In other words, the block format 
is preserved by non-singular matrices $F$ which are even 
in the sense of block format matrices (and only by these). 

Henceforth, we define a {\em matrix format}
to be a choice of ordered homogeneous basis of the 
graded vector space $V$, modulo even changes of basis.
For each format, the involution matrix is diagonal 
with entries $\pm 1$ and an even change of basis leaves this matrix 
invariant (i.e. $\epsilon ^{\prime} = \epsilon$). 

A generic change of basis (e.g. leading to 
a non homogeneous basis) leads to another matrix realization, 
but in general not to another matrix format.   
In the following, we will study {\em format changing transformations}, 
i.e. changes of basis which relate 
different matrix formats. They simply permute 
the diagonal elements of the involution matrix.

\section{Passage from the block format of 
$gl(n|m)$ to other formats}

Starting from the block format, 
we consider a different ordering for the homogeneous basis 
$\{ e_1, ..., e_n \} \bigcup \{ e_{n+1} ,..., e_{n+m} \}$
of $V$:  
instead of arranging the even and odd components
of a vector into blocks,
we arrange them in a
different way by considering a permutation
$P$ of the ordered set $(1,...,n+m)$ which mixes 
$(1,...,n)$ and $(n+1,...,n+m)$:
in matrix notation,
this linear transformation of vectors reads as 
\begin{equation}
\label{f}
v = \left[ v_1, v_2,... ,v_{n+m} \right]^T
\ \stackrel{F}{\longmapsto} \
v^{\prime} = \left[ v_{P(1)},  v_{P(2)},... ,v_{P(n+m)} \right]^T
\end{equation}
or 
\begin{equation}
\label{f1}
v^{\prime} = Fv \qquad {\rm with} \qquad
F_{ij} = \de_{P(i)j}
\ \ .
\end{equation}
The mapping $F$ (which satisfies $F^{-1}=F^T$)
represents the simplest example of a format changing transformation. 
It induces  the following 
similarity transformation 
for the matrices (\ref{bl}) :
\begin{equation}
\label{simt}
M^{\prime}  =  FMF^{-1} \qquad {\rm i.e.} \qquad
M^{\prime}_{ij} = M_{P(i)P(j)}
\ \ .
\nonumber
\end{equation}
The involution $\epsilon ^{\prime}  =  F\epsilon F^{-1}$ differs
from $\epsilon$ and the grading of $M^{\pr}$ is 
related to the one of $M$ by
\begin{equation}
\label{gra}
\dd \, M^{\pr}_{ij}  =
\dd \, M_{P(i)P(j)} 
\ \ .
\end{equation}
Since the trace is coordinate-independent, we have 
${\rm tr} \, \epsilon M
= {\rm tr} \, \epsilon^{\pr} M^{\pr}$, henceforth 
\begin{equation}
\label{st}
{\rm str} \, M^{\pr} =
{\rm tr} \, \epsilon^{\pr} M^{\pr}
\ \ .
\end{equation}
According to equation (\ref{simt}),
the principal diagonal elements
remain principal diagonal elements and therefore this 
expression still 
represents a sum of such elements with certain signs.
 
The supercommutator 
of the matrices $M^{\pr}$ and $N^{\pr}$ is related to the one 
of $M$ and $N$ by 
\begin{equation}
\label{ssc}
[ M^{\pr} , N^{\pr} \} :=
F [ M , N \} F^{-1}
\end{equation}
and the 
supertranspose of $M^{\pr}$ is defined by
\begin{equation}
\label{tt}
\left( M^{\pr} \right) ^{sT} := F M^{sT} F^{-1} 
\ \ .
\end{equation}
 
Just as for matrices in block format, 
the format of the matrices $M^{\prime}$  
is preserved by 
non-singular endomorphisms
which are even in the sense of the new format.

\section{$gl(n + 1 |n)$ in diagonal format}

Consider 
${\rm dim} \, V_{\0} =n+1$ and ${\rm dim} \, V_{\1} =n$
with $n\geq 1$. 
The even and odd components
of a column vector 
$[v_1,...,v_{2n+1}]^T$ in block format 
can be rearranged in an
alternating way (even/odd/even/odd/...). 
The {\em simplest} such arrangement is the one which respects 
the order :  
\begin{equation}
\label{simpl}
v = \left[ v_1,...,v_{n+1} , v_{n+2},...,v_{2n+1} \right]^T
\ \stackrel{F}{\longmapsto} \
v^{\prime} = \left[ v_1,v_{n+2}, v_{2},v_{n+3},...,
v_{2n+1},v_{n+1} \right]^T
\ .
\end{equation}
In terms of the notation (\ref{f})(\ref{f1}), 
this transformation
corresponds to the permutation
\begin{eqnarray*}
P(2i+1) & = & i+1 \qquad \quad \ \ \,
{\rm for} \quad 0 \leq i \leq n
\\
P(2i) & = & n+i+1 \qquad {\rm for} \quad 1 \leq i \leq n
\ \ .
\end{eqnarray*}
The matrix 
$M^{\prime}$ given by eq.(\ref{simt}) then represents a 
$(2n+1)\times (2n+1)$ 
matrix with alternating even and odd diagonals
(alternating even and odd elements in the rows and columns).
The different diagonals are labeled by $i$ 
where $-n \leq i \leq n$ and where the main diagonal is
counted as the $0$-th one; 
a matrix which has entries 
$a_1, a_2,...$ along the $i$-th diagonal,
all other entries being zero, is denoted 
by ${\rm diag}_i (a_1, a_2,...)$.
We will say for short that such a matrix only has entries    
on the  $i$-th diagonal. 
We have the explicit expressions
\begin{eqnarray}
\dd \, M^{\pr}_{ij} & = &  i+j \  {\rm mod} \, 2
\nonumber \\
\epsilon^{\prime} & = &
{\rm diag}_0 \, \left( 1,-1,1,-1,...,1 \right)
\nonumber \\
{\rm str} \, M^{\pr} & = &
\sum_{i=1}^{2n+1} \, (-1)^{i+1} \,
M^{\pr}_{ii}
\label{di}
\\
{[ M^{\pr} , N^{\pr} \} _{ik}}
& = &
\sum_{j=1}^{2n+1} \left( M^{\pr} _{ij} N^{\pr} _{jk}
\, - \, (-1)^{(i+j)(j+k)} \,
N^{\pr} _{ij} M^{\pr} _{jk} \right)
\nonumber \\
\left( M^{\pr \, sT} \right) _{ij} & = &
(-1)^{i(i+j)} M^{\pr} _{ji}
\ \ .
\nonumber
\end{eqnarray}
This format of representative column vectors and matrices 
will be called the 
{\em diagonal format}.
Thus, $gl(n+1|n)$ in diagonal format is the Lie superalgebra
of complex $(2n+1)\times (2n+1)$ matrices with a grading,
super-commutator, -trace and -transpose defined by
eqs.(\ref{di}).
The non-singular matrices all of whose odd diagonals vanish 
are those endomorphisms which preserve the diagonal format - 
see next section for examples. 

Note that these expressions (\ref{di}) 
are easier to manipulate than those for the block format 
since the sign factors 
refer directly to the indices rather than to their parity,
e.g. 
\[
\epsilon^{\pr}_{ij} = (-1)^{i+1} \delta_{ij} 
\quad , \quad 
\left( ( M^{\pr} )^{sT} \right) ^{sT} _{ij} =
(-1)^{i+j} M^{\pr} _{ij}
= \left( {\rm Ad}_{\epsilon ^{\pr} }\, M^{\pr} \right) _{ij}
\ \ .
\]

Obviously, the alternating partitioning can be performed
in a similar way in the case of the Lie superalgebras 
$gl(n|n+1)$ or $gl(n|n)$.

\section{Fermionic superalgebras in diagonal format}

In this section, we discuss the diagonal format of different 
subalgebras of $gl(n+1|n)$, namely of $sl(n+1|n)$
and of the orthosymplectic algebras $osp(2m\pm 1|2m)$
with $4m\pm 1 = 2n+1$. We do so by elaborating on the expressions
encountered in the study of 
$W$ superalgebras
in reference \cite{gt}.
   
\subsection{$sl(n+1|n)$}
 
A general element of $sl(n+1|n)$ in diagonal format is 
simply a $(2n+1)\times(2n+1)$ matrix of numbers together with
an even (odd) grading for the even (odd) diagonals,
the only restriction being the vanishing of the supertrace.

The Cartan matrix elements $a_{ij}$
of $sl(n+1|n)$ 
are independent of the chosen format and define a $2n\times 2n$ 
matrix.
If all simple roots are chosen to be fermionic, this matrix reads as  
\cite{japan}
$$
(a_{ij}) = 
\sum _{k=\pm 1} {\rm diag}_k (1,-1,1-1, \dots ) 
$$
and its inverse  
$$
(a^{ij}) = 
\sum _{k=\pm 1, \pm 3,...}
{\rm diag}_k (1,0,1,0, \dots ) 
$$
has the property 
\begin{equation}
\label{cars}
\sum_j a^{ij} = \left\{ 
\begin{array}{ll}
{i \over 2} & \qquad {\rm for} \ \; i \ \; {\rm even} \\
n - {i-1 \over 2} & \qquad {\rm for} \ \; i \ \; {\rm odd} \ \ .
\end{array} 
\right.  
\end{equation}

Let $E_{ij}$ denote the matrix with elements
$\left( E_{ij} \right) _{kl} \, = \,
\de_{ik} \de_{jl}$. 
As Cartan and Chevalley generators of $sl(n+1|n)$,  
we can take the 
matrices \cite{gt} $(i= 1, \dots , 2n )$ 
\begin{eqnarray}
h_i & = & (-1)^{i+1} \, \left( E_{ii} + E_{i+1,i+1} \right) 
\nonumber 
\\
\label{slch}
e_i & = & (-1)^{i+1} \, E_{i, i+1}
\\
f_i & = & E_{i+1, i}
\ \ ,
\nonumber 
\end{eqnarray}
i.e. the Cartan generator 
$h_i \equiv \{ e_i , f_i \}$ 
only has entries on the main diagonal
while $e_i$ and $f_i$ only have entries on the first upper and lower
diagonals, respectively. 

According to equations (\ref{spal}) and (\ref{cars}), 
the principal embedding 
$osp(1|2)_{\rm pal}\subset sl(n+1|n)$
is generated by the $(2n+1) \times (2n+1)$ matrices
\begin{eqnarray}
J_- &=&  {\rm diag} _{-1}(1,\dots,1)
\nonumber  
\\
J_+ &=& {\rm diag} _{+1}(n,-1,n-1,-2,\dots,-n)
\label{sppal}
\\
H &=& {\rm diag} _0 (n,n-1,\dots,-n) 
\ \ , 
\nonumber 
\end{eqnarray}
which are, respectively, symmetric and antisymmetric 
with respect to their antidiagonal.

\paragraph{Highest weights}

The {\em highest weight generators} of $osp(1|2)_{\rm pal}$
are defined as those $M_k\in sl(n+1|n)$ 
(with $k=1, \dots ,2n$) which satisfy
\begin{equation}
\label{high}
{[H,M_k ]} = k M_k  
\qquad {\rm and} \qquad
{ [ J_+,M_k \} } = 0 
\ \ .
\end{equation}
For the solutions of these conditions one finds $M_k=M_1^k$ with
\begin{equation}
\label{hweights}
M_1 = {\rm diag} _{+1}(n,1,n-1,2,\dots,n)
\qquad \quad 
(\ M_1^{2n+1}=0\ )
\ \ .
\end{equation} 
Thus, $M_k$ is a matrix which only has entries on the $k$-th diagonal 
and which is symmetric 
with respect to the antidiagonal.

\subsection{The subalgebras $osp(2m\pm 1|2m)$}

Consider the graded algebras $osp(2m\pm1|2m)\subset sl(n+1|n)$
with $4m\pm1=2n+1$. 
Their rank is 
$2m$ for $osp(2m+1|2m)$ and $2m-1$ for $osp(2m-1|2m)$.
 
By definition \cite{cor},
the superalgebra $osp(2m\pm1|2m)$ consists of
supermatrices $M$ which satisfy $M^{sT}G+GM=0$
where the supermetric $G=G_{so(2m\pm1)}\oplus G_{sp(2m)}$
is a direct sum of $so(2m\pm1)$ and $sp(2m)$ metrics.  
The latter represent symmetric, respectively
antisymmetric, non-degenerate bilinear forms 
on a complex vector space.

In a basis of the root space with only {\em fermionic 
simple roots},
the algebras
$osp(2m\pm1|2m)$ are characterized by Cartan matrices
 whose only non-zero elements 
are \cite{japan}
\begin{equation}
a_{11}=1 \ \  , \  \ a_{i,i+1}=a_{i+1,i}=(-1)^i 
\quad {\rm with} \, \left\{
\begin{array}{ll}
i = 1,.., 2m-2  & \  {\rm for} \  osp(2m-1|2m) 
\\
i = 1,..,2m-1 &\ {\rm for} \   osp(2m+1|2m).
\end{array}
\right.
\end{equation}
The inverse matrix for $osp(2m-1|2m)$ is given by  
$$
(a^{ij}) = 
\sum _{k= 0, \pm 2,...}
{\rm diag}_k (1,0,1,0, \dots )
+
\sum _{k=\pm 1, \pm 3,...}
{\rm diag}_k (0,1 ,0,1, \dots ) 
$$ 
and for $osp(2m+1|2m)$ it reads  
$$
(a^{ij}) =  \sum _{k= 0, \pm 2,...}
{\rm diag}_k (0,-1,0,-1, \dots )
+
\sum _{k=\pm 1, \pm 3,...}
{\rm diag}_k (-1 ,0,-1,0, \dots ) 
\ \ .  
$$

We will now discuss the two cases $osp(2m-1|2m)$ and
$osp(2m+1|2m)$ in turn. Their block format is summarized 
in appendix A.
The transition between matrices in block and 
diagonal format is mediated
by a similarity transformation,
$M_{\rm diag.}=L^{-1}M_{\rm block}L$.
We will not choose the matrix $L$ 
according to the simplest alternating arrangement 
as given by equation (\ref{simpl}), but rather according to 
a more complicated arrangement which leads to diagonal 
format and which allows us to recover the 
expression (\ref{sppal}) for the
generators of $osp(1|2)_{\rm pal}$.
This requirement determines the form of the Chevalley 
generators of $osp(2m\pm 1|2m)$, see below. 
By expanding with respect 
to the corresponding Chevalley basis, it follows that 
a general $osp(2m\pm 1|2m)$ element in diagonal format is 
a $(4m\pm 1)\times(4m\pm 1)$ matrix for which the 
even and odd elements satisfy 
the following conditions, respectively:
\begin{eqnarray}
M_{i,i+2k} &=& (-1)^{k+1} \, M_{p+1-i-2k,p+1-i}
\nonumber
\\
\label{ospelement}
M_{i,i+2k+1}&=& (-1 )^{k+1} \, M_{p-i-2k,p+1-i}
\\
(\ p=4m\pm 1  \   &,&  \  i\in \{1,\dots , p \} \quad , \quad
k \in \{ 0, \pm 1,\dots \} \ ) 
\ \ .
\nonumber 
\end{eqnarray}
Thus, the even diagonals of $M$ are alternatingly antisymmetric 
and symmetric with respect 
to the antidiagonal and so are the odd diagonals, 
e.g. see expression (\ref{1}) for 
$M \in osp(3|2)$.

\subsubsection{$osp(2m-1|2m)$}

The choice of $L$ that leads to the expression (\ref{sppal})
for the
generators of $osp(1|2)_{\rm pal}$ is 
\begin{eqnarray}
\label{sim}
L &=& \sum_{i=0}^{m-1}(-1)^{i+1}E_{2m+i,2i+1}
+\sum_{i=1}^m(-1)^i
E_{2m-i,2i}
\\
&& \qquad \qquad +(-1)^{m+1}\sum_{i=1}^m E_{4m-i,2m+2i-1}
+(-1)^m \sum_{i=1}^{m-1}E_{m-i,2m+2i}
\ \ ,
\nonumber
\end{eqnarray}
which  matrix satisfies $L^{-1} = L^{T}$. This endomorphism 
describes a permutation of basis vectors together with some changes 
of signs and thus defines a format changing transformation followed by 
a format preserving endomorphism. 
It leads to diagonal format with an 
involution given by $\epsilon_{ij} = (-1)^i \delta_{ij}$, i.e. 
the diagonal format matrices have 
even and odd elements on the even and odd diagonals, respectively. 

The Chevalley generators can be represented as $(i=1, \dots , 2m-1)$
\begin{eqnarray}
h_i &=& (-1)^{i+1}
\Bigl(E_{2m-i,2m-i}-E_{2m+i,2m+i}
    +E_{2m+1-i,2m+1-i}-E_{2m-1+i,2m-1+i}\Bigr)
\nonumber
\\
\label{chevalleym}
e_i &=& (-1)^i
\Bigl(E_{2m-1+i,2m+i}-E_{2m-i,2m+1-i}\Bigr)
\\
f_i &=& E_{2m+i,2m-1+i}+E_{2m+1-i,2m-i}
\nonumber
\end{eqnarray}
i.e. the Cartan generators only have entries on the main diagonal
while $e_i$ ($f_i$) only has entries on the first upper (lower) 
diagonal.

The metric matrix takes the form 
\begin{equation}
\label{inv}
G \; = \; (-1)^{m} \, 
\sum_{i=1}^{4m-1}(-1)^{\lbrack{i+1\over2}\rbrack}
E_{i,4m-i} 
\; = \; (-1)^{m} \,
{\rm adiag} \, (1,-1,-1,1, 1,- 1, -1,...)  
,
\end{equation}
where $\lbrack i\rbrack$ denotes the integer part of $i$ 
and where ${\rm adiag} \, (a_1, a_2,...)$ represents a matrix which 
only has entries $a_1,a_2,...$ on the antidiagonal.  
The fact that the factors $1$ and $-1$ occur by pairs 
in the metric (\ref{inv}), reflects itself in the matrices 
(\ref{ospelement}) of $osp(2m - 1 |2m)$ by the fact that 
the symmetric and antisymmetric diagonals occur by pairs
(cf. eq.(\ref{1}) for the $osp(2m + 1 |2m)$ elements 
which have the same symmetry properties).

For the even and odd elements
of $M^{sT}$, we find 
\begin{eqnarray}
\left(M^{sT}\right)_{i,i+2k} &=& M_{i+2k,i} 
\nonumber 
\\
\label{supertrans}
\left(M^{sT}\right)_{i,i+2k+1} &=& (-1)^{i} \, M_{i+2k+1,i}
\\
(\ i\in \{1,\dots , 4m-1 \} & \ , \ & 
k \in \{ 0, \pm 1,\dots \} \ )
\ \ ,
\nonumber
\end{eqnarray} 
i.e.  
supertransposition coincides with ordinary transposition 
apart from the fact that one also has to change the 
signs on all odd diagonals
in an alternating way.

\subsubsection{$osp(2m+1|2m)$}

The results have the same form and characteristics 
as those for $osp(2m-1|2m)$
and therefore we will only list the expressions. 

\noindent - Format changing matrix $(L^{-1} = L^{T})$ :  
\begin{eqnarray}
L&=& \sum_{i=0}^{m}(-1)^{i}E_{2m+1-i,2i+1}+\sum_{i=1}^m
\left[ (-1)^i
E_{2m+1+i,2i} \right.
\label{simt1}
\\
&& \left.
\qquad\qquad+(-1)^{m+1}  E_{4m+2-i,2m+2i}
+(-1)^m E_{m+1-i,2m+1+2i} \right] 
\nonumber 
\end{eqnarray}
\noindent  - Involution :
\begin{equation} 
\epsilon_{ij} = (-1)^{i+1} \delta_{ij}
\end{equation}   

\noindent - Cartan-Chevalley generators $(i=1, \dots , 2m)$ : 
\begin{eqnarray}
h_i &=& (-1)^{i+1}
\Bigl(E_{2m+1-i,2m+1-i}-E_{2m+1+i,2m+1+i}
    +E_{2m+2-i,2m+2-i}-E_{2m+i,2m+i}\Bigr)
\nonumber 
\\
\label{chevalleyp}
e_i &=& (-1)^i
\Bigl(E_{2m+i,2m+1+i}-E_{2m+1-i,2m+2-i}\Bigr)
\\
f_i &=& 
E_{2m+1+i,2m+i}+E_{2m+2-i,2m+1-i}
\nonumber 
\end{eqnarray}

\noindent - Supermetric : 
\begin{equation}
\label{metr}
G \; = \; (-1)^{m} \, 
\sum_{i=1}^{4m+1}(-1)^{\lbrack{i\over2}\rbrack}
E_{i,4m+2-i}
\; = \; (-1)^{m} \, 
{\rm adiag} \, (1, 1,-1,-1,1, 1,...)  
\end{equation}

\noindent - Supertranspose : 
\begin{eqnarray}
\label{supertrans1}
\left( M^{sT} \right) _{i,i+2k} & = & 
M_{i+2k,i} \quad  \quad
\nonumber
\\
\left( M^{sT} \right) _{i,i+2k+1} &=& 
(-1)^{i+1} \, M_{i+2k+1,i} 
\\
(\  i\in \{1,\dots , 4m+1 \}  & \  , \ & 
k\in \{ 0, \pm 1,\dots \}\ ) \ \ .
\nonumber
\end{eqnarray}

\paragraph{Principal $osp(1|2)$-embedding and highest weights}

The basis (\ref{chevalleym}) and (\ref{chevalleyp})
have been chosen such as to lead
to the same explicit expressions for 
the principal embedding  
$osp(1|2)_{{\rm pal}} \subset osp(2m\pm1|2m)$
as in the $sl(n+1|n)$ case, see
equation (\ref{sppal}). 
The highest weight generators are also the
same, except that we have to exclude
those $M_k$ which do not belong to
$osp(2m\pm1|2m)$. This leaves us with 
$M_{2+4q}$ and 
$M_{3+4q}$ where $q=0,1,\dots$

\subsection{Infinite dimensional limit}

For $n \to \infty$ and $m \to \infty$, the algebras 
$sl(n+1|n)$ and $osp(2m\pm1|2m)$ of diagonal format matrices 
give rise to infinite dimensional Lie superalgebras  
denoted by $sl_{\infty}$ and $osp_{\infty}$, 
respectively. Similar (and 
closely related) algebras occur in the study 
of superintegrable models (KP-hierarchy,...) and 
have been introduced and discussed 
in references \cite{kac}.
  
In order to derive infinite dimensional algebras  
from our expressions, 
we label the matrix entries $M_{ij}$ by $i,j \in {\bf Z}$ 
and consider the diagonal format grading defined by 
${\rm deg} \, M_{ij} =i+j$.
Then, expressions (\ref{slch}) with $i \in {\bf Z}$ 
represent a Chevalley basis 
of $sl_{\infty}$. A basis of 
$osp_{\infty}$ is given by the matrices
\begin{equation}
E_{i,j} - (-1) ^{ ] {1 \over 2} (i-j) [ } \, E_{-j , -i}
\ \ ,
\end{equation}
where $]k[$ denotes the smallest integer greater or equal to $k$.
(This result simply amounts to enlarging {\em ad infinitum} the matrix
$M_{ {\rm diag.} }$ of eq.(\ref{1}).)
A Chevalley basis of $osp_{\infty}$ reads $(i=0,1,2,...)$
\begin{eqnarray}
e_i &=& E_{i,i+1} - E_{-i-1,-i} 
\nonumber 
\\
f_i &=& E_{i+1,i} + E_{-i,-i-1}
\\
h_i &=& E_{i+1,i+1} 
+ E_{i,i} - E_{-i,-i}
- E_{-i-1,-i-1} 
\ \ . 
\nonumber 
\end{eqnarray}   
These results are equivalent to those given and applied in references
\cite{kac}.

\section{Concluding remarks}

In the previous section, 
we have not discussed all the 
fermionic Lie superalgebras listed in eq.(\ref{class}). 
For $osp(2m|2m)$, the diagonal format can be considered, but due 
to the fact that $2m+2m \neq 2n+1$,   
it is not possible to recover the expressions (\ref{sppal}) 
for the generators of $osp(1|2)_{ {\rm pal} } \subset 
sl(n+1|n)$  
(as one might wish for the applications to $W$-superalgebras).  
For $osp(2m+2|2m)$, alternating formats can also be introduced 
for vectors and matrices; the simplest and most symmetric 
such arrangement is obtained by putting the two extra 
even entries in the middle, e.g. for the involution of 
$osp(6|4)$: 
\[
\epsilon = {\rm diag} _0 \, ( 1,-1,1,-1,1,1,-1,1, -1,1)
\ \ .
\]
The matrices of $osp(2m+2|2m)$ in this format then have a symmetry 
structure which is quite similar
to the one for the diagonal format of $osp(2m\pm 1|2m)$.

We note that one can also introduce the 
{\em superprincipal embedding of} $sl(2|1)$ in 
$sl(n+1|n)$ and derive from it the one of $osp(1|2)$:
$osp(1|2)_{ {\rm pal} } \subset sl(2|1)_{ {\rm pal} } 
\subset sl(n+1|n)$.
Explicit formulas have been given in references 
\cite{drs, fss} and diagonal format expressions have been 
applied to superintegrable models 
and $W$-algebras in \cite{dm,gg}.

Among the various inequivalent simple root systems (SRS's)
of a basic Lie superalgebra, there exists a canonical one, 
the so-called {\em distinguished SRS} \cite{fss,vk}: it is 
characterized by the fact that it contains the smallest possible 
number of odd roots,  
e.g. for $sl(n+1|n)$, this system only contains one odd root. 
The so-called {\em fermionic SRS} (which exists for the Lie superalgebras 
listed in eq.(\ref{class}) and which 
has been considered in the present paper) 
represents the other extreme 
where {\em all} simple roots are chosen to be odd.
Both SRS's can be related by a generalized Weyl transformation 
\cite{fss}.
In order to get a better understanding of the graded structure of the 
root space of $sl(n+1|n)$ (which is of dimension $2n$), 
we consider  $sl(3|2)$ as an example. 
To start with, we do not specify the matrix format and we 
assume that the generators $e_i$ associated to a SRS only have entries on  
the first upper diagonal, i.e.  
$e_i = E_{i,i+1}$ with $i=1,2,3,4$. 
If the block matrix format is considered, one of the generators
is odd  
(namely $e_3 = E_{3,4}$) while all the others are even. 
A generic element of the 
Cartan subalgebra is then parametrized by    
$h_{ {\rm block} } = {\rm diag}_0 \, (\varepsilon _1, \varepsilon _2, 
\varepsilon _3, \delta_1, \delta_2)$ 
with 
$\varepsilon _1 + \varepsilon _2 + \varepsilon _3 = \delta_1 +\delta_2$
and from the eigenvalue equation 
\begin{equation}
\label{ev}
[ h , e_i ] = \alpha_i  e_i  
\end{equation}
we obtain the distinguished SRS of $sl(3|2)$: 
$$
\alpha_1 = \varepsilon _1 - \varepsilon _2 
\quad , \quad 
\alpha_2  = \varepsilon _2 - \varepsilon _3
\quad , \quad 
\alpha_3  = \varepsilon _3 - \delta_1 
\quad , \quad 
\alpha_4 = \delta_1 - \delta_2  
\ \ .
$$ 
On the other hand, if the diagonal matrix format is chosen, 
the generators $e_i$ are all odd. The  
Cartan subalgebra is then parametrized by    
$h_{ {\rm diag.} } = {\rm diag}_0 \, (\varepsilon _1, \delta_1,
\varepsilon _2, \delta_2 , 
\varepsilon _3 )$ 
with 
$\varepsilon _1 + \varepsilon _2 + \varepsilon _3 = \delta_1 +\delta_2$
and it follows from eq.(\ref{ev}) that the fermionic SRS 
of $sl(3|2)$ reads as  
$$
\alpha_1  = \varepsilon _1 - \delta_1  
\quad , \quad 
\alpha_2  = \delta_1 - \varepsilon _2 
\quad , \quad 
\alpha_3  = \varepsilon _2 - \delta_2 
\quad , \quad 
\alpha_4  = \delta_2 - \varepsilon _3  
\ \ .
$$ 
From this point of view, the number of odd roots in a SRS of 
$sl(n+1|n)$ (and more generally of $sl(m|n)$) simply coincides
with the number of sign changes in the involution defining the format
under consideration, 
e.g. for the block and diagonal formats of 
$sl(3|2)$, we respectively have one and four such changes of signs: 
\[
M_{{\rm block}} = \left[
\begin{array}{ccccc}
+ &    &    &    &     \\
  & +  &    &    &     \\
  &    & +  &  \bullet  &     \\
  &    &    & -  &     \\
  &    &    &    &  -    
\end{array}
\right] 
\ \ , \ \ 
M_{{\rm diag.}} = \left[
\begin{array}{ccccc}
+ & \bullet   &    &    &     \\
  & -  & \bullet   &    &     \\
  &    & +  &  \bullet  &     \\
  &    &    & -  & \bullet    \\
  &    &    &    &  +    
\end{array}
\right] 
\ \ .
\]
This argument also explains why it is not possible to 
find a fermionic SRS
for $sl(m|n)$ for $m\neq n+1$. 

For the sake of clarity, we should emphasize that the choice of 
a specific SRS 
does not impose or imply a choice of matrix format,  
e.g. once one has chosen the fermionic SRS 
and the diagonal matrix format, one can 
go over  to the block (or any other) matrix  
format by virtue of a similarity transformation, 
the consequence being that the 
root generators $e_i$ loose their simple 
form (of matrices with entries on the first upper diagonal only).   
The previous discussion 
confirms that 
the diagonal matrix format (for which there
is a {\em maximal} number of even/odd alternations in column vectors 
and matrices) is best adapted to 
fermionic SRS's  
while the block format (for which there is {\em no} even/odd alternation 
at all) is best suited for the distinguished SRS.  
For intermediate SRS's, an intermediate 
matrix format is appropriate and can 
be introduced along the lines described in the text.

In summary, we have studied the general structure of specific 
supermatrix arrangements which have previously 
been encountered in the literature. We hope that our 
systematic presentation elucidates these examples  
and that it will prove to be useful for further applications 
in physics and mathematics.

\vskip 1.5truecm
 
{\bf \Large Acknowledgments}
 
\vspace{3mm}
 
F.G. wishes to thank D. Maison for the hospitality 
extended to him at 
the Max-Planck-Institut 
and M. Schottenloher for his invitation 
to lecture on supersymmetry at the Graduiertenkolleg 
of the LMU Munich.

\newpage

\setcounter{equation}{0}
\renewcommand{\theequation}{A.\arabic{equation}}  

\appendix
\section{$osp$-superalgebras in block format}

If the standard matrix format is chosen for $osp(2m\pm 1|2m)$,  
the supertranspose of $M$ is defined by
equation (\ref{sst}) and the basis of $V_{\0}$ and $V_{\1}$ 
can be chosen in such a way that 
the metrics are given by 
\begin{equation}
\label{supm}
G_{so(2m\pm1)}=
\left[
\begin{array}{cccc}
 & & &1 \\
 & & \cdot & \\ 
 & \cdot & & \\
1 & & &  
\end{array} \right]
\qquad , \quad
G_{sp(2m)}=
\left[
\begin{array}{cc}
0&-{\bf 1}_m 
\\
{\bf 1}_m & 0 
\end{array} \right] 
\ \ .
\end{equation}
This form implies that 
the matrices belonging to  
$so(2m\pm1)$ are antisymmetric with respect to the antidiagonal 
(i.e. $A_{ij} = - A_{p+1-j, p+1-i}$ with $p=2m\pm 1$) 
and that 
the Cartan subalgebra
of $osp(2m\pm 1|2m)$
consists of diagonal matrices.

\paragraph{\underline{$osp(2m-1|2m)$}}

\bigskip 

Among the independent matrix entries, 
there are $4m^2-2m+1$ even and $2m(2m-1)$ odd elements. 
The
Chevalley basis in block format is represented by the matrices 
\cite{gt}
\begin{eqnarray}
e_{2i-1} &=& E_{m+1-i,4m-i}+E_{3m-i,m-1+i}
\quad \qquad (\ i = 1,\dots,m \ )  
\nonumber
\\
e_{2i} &=& E_{m+i,3m-i}-E_{4m-i,m-i}
\quad\qquad\qquad \! (\ i = 1,\dots,m-1 \ )  
\nonumber 
\\
&&
\nonumber
\\
\noalign{\vskip.3cm}
f_{2i-1} &=& E_{m-1+i,3m-i}-E_{4m-i,m+1-i}
\nonumber
\\
f_{2i} &=& -E_{m-i,4m-i}-E_{3m-i,m+i}
\label{bc}
\\
&&
\nonumber
\\
\noalign{\vskip.3cm}
h_{2i-1} &=& E_{3m-i,3m-i}-E_{4m-i,4m-i}-E_{m+1-i,m+1-i}+E_{m-1+i,m-1+i}\cr
h_{2i} &=& E_{m-i,m-i}-E_{m+i,m+i}-E_{3m-i,3m-i}+E_{4m-i,4m-i}
\ \ .
\nonumber
\end{eqnarray}

\paragraph{\underline{$osp(2m+1|2m)$}}

\bigskip 

There are $2m(2m+1)$ even and as many odd independent matrix elements.
The Chevalley basis in block format is \cite{gt} 
($i =1,\dots, m$)
\begin{eqnarray}
e_{2i-1} &=& E_{m+2-i,4m+2-i}+E_{3m+2-i,m+i}
\nonumber
\\
e_{2i} &=&  E_{m+1+i,3m+2-i}-E_{4m+2-i,m+1-i}
\nonumber
\\
&& 
\nonumber
\\
f_{2i-1} &=& E_{m+i,3m+2-i}-E_{4m+2-i,m+2-i}
\nonumber
\\
f_{2i} &=& -E_{m+1-i,4m+2-i}-E_{3m+2-i,m+1+i}
\label{horse}
\\
&& 
\nonumber
\\
h_{2i-1} &=& E_{3m+2-i,3m+2-i}-E_{4m+2-i,4m+2-i}
-E_{m+2-i,m+2-i}+E_{m+i,m+i}
\nonumber
\\
h_{2i} &=& E_{m+1-i,m+1-i}-E_{m+1+i,m+1+i}
-E_{3m+2-i,3m+2-i}+E_{4m+2-i,4m+2-i }
\ \  .
\nonumber
\end{eqnarray}


\newpage
 
\begin{thebibliography}{22}
\newcommand{\artref}[5]{{\sc #1}: {\it #2}, #3 {\bf #4} #5}
\newcommand{\bookref}[3]{{\sc #1}: ``{\it #2}$\,$", #3}
\newcommand{\prepref}[3]{{\sc #1}: {\it #2}, #3}
 
 
\bibitem{vk}
\artref
{V.G.Kac}{Lie superalgebras}{
Adv.Math.}{26}{(1977) 8-96 ;}
 
\artref
{V.G.Kac}{A sketch of Lie superalgebra theory}{
Commun.Math.Phys.}{53}{(1977)
31-64 .}
 
\bibitem{cns}
\artref
{L.Corwin, Y.Ne'eman and S.Sternberg}{Graded Lie algebras
in mathe\-matics
and physics (Bose-Fermi symmetry)}{Rev.Mod.Phys.}{47}{(1975) 573-603 .}
 
\bibitem{scheu}
\bookref
{M.Scheunert}{The theory of Lie superalgebras}{Lecture Notes
in Mathe\-matics 716 (Springer, Berlin 1979) .}
 
\bibitem{ber}
\bookref
{F.A.Berezin}{Introduction to Superanalysis}{Mathematical
Physics and Applied Mathematics Vol.9
(D.Reidel Publ. Co., Dordrecht 1987) .}
 
\bibitem{man}
\bookref
{Yu.I.Manin}{Gauge field theory and complex geometry}{
Grundlehren der mathematischen Wissenschaften Vol.289
(Springer, Berlin 1988) .}

\bibitem{cons}
\bookref
{F.Constantinescu and H.F. de Groote}{Geometrische und 
algebrai\-sche Methoden der Physik: Supermannigfaltigkeiten 
und Virasoro-Algebren}{(Teubner, Stuttgart 1994) .}

\bibitem{bdw}
\bookref
{B.deWitt}{Supermanifolds}{(Cambridge University Press, Cambridge 1984) .}
 
\bibitem{cor}
\bookref
{J.F.Cornwell}{Group Theory in Physics Vol.3: Supersymmetries and
infinite-dimensional algebras}{
(Academic Press, London 1990) .}

\bibitem{fss}
\bookref
{L.Frappat, A.Sciarrino and P.Sorba}{Dictionary on Group Theory:
Superalgebras}{
ENSLAPP-AL-600/96, hep-th/9607161 .} 

\bibitem{gt}
\artref
{F.Gieres and S.Theisen}{Superconformally covariant operators
and super $W$ algebras}{J.Math.Phys.}{34}{(1993) 5964-5985 ;}
 
\artref
{F.Gieres and S.Theisen}{Classical $N=1$ and $N=2$ super $W$-algebras
from a zero-curvature
condition}{Int.J.Mod.Phys.}{A9}{(1994) .}
 
 \bibitem{dm}
\artref
{F.Delduc and M.Magro}{Gauge invariant formulation of $N=2$ Toda and KdV systems in extended 
superspace}{J.Phys.}{A29}{(1996) 4987-5000 .}
 
\bibitem{gg}
\artref
{F.Gieres and S. Gourmelen}{$d=2, N=2$ superconformally covariant
operators and super $W$-algebras}{J.Math.Phys.}{39}{(1998)
3453-3475 .}

\bibitem{bsa}
\artref
{L.Benoit and Y.Saint-Aubin}{Singular vectors of the Neveu-Schwarz 
algebra}{Lett.Math.Phys.}{23}{(1991) 117-120 .}
 
\bibitem{wl}
\artref
{B.Bershadsky, W.Lerche, D.Nemeschansky and N.P.Warner}{Extended
$N=2$ superconformal structure of gravity and $W$-gravity 
coupled to matter}{Nucl.Phys.}{B401}{(1993) 304-347 .}
  
\bibitem{qg}  
\bookref
  {P.P.Kulish}{Quantum superalgebra $osp(2|1)$}{preprint 
  RIMS-615, Kyoto 1988 ;} 

\artref
{P.P.Kulish and N.Yu.Reshetikhin}{Universal $R$-matrix of the quantum
superalgebra $osp(2|1)$}{Lett.Math.Phys.}{18}{(1989) 143-149 ;} 

\artref
{H.Saleur}{Quantum  $osp(1|2)$ and solutions of the graded Yang-Baxter
equation}{Nucl.Phys.}{B336}{(1990) 363-376 ;} 

\artref
{E.Celeghini and P.P.Kulish}{Twist deformation of the rank-one Lie 
superalgebra}{J.Phys.A: Math.Gen.}{31}{(1998) L79-L84 .} 
  
 
\bibitem{kac}
\artref
{V.G.Kac and J.W. van de Leur}{Super boson-fermion 
correspondence}{Ann.Inst.Fourier}{37}{(1987) 99-137 ;}

\bookref
{V.G.Kac and J.W. van de Leur}{Super boson-fermion 
correspondence of type B}{in ``Infinite dimensional Lie algebras 
and groups", V.G.Kac, ed. (World Scientific, 1989), pp.369-406 .} 

\bibitem{lss}
\bookref
{D.A.Leites, M.V.Saveliev and V.V.Serganova}{Embeddings of 
$osp(N|2)$ and the associated nonlinear supersymmetric equations}{in 
`Group Theoretical 
Methods in Physics Vol.1', M.A.Markov, V.I.Manko  and V.V.Dodonov, eds. 
(VNU Science Press, Utrecht 1986) .}
 
\bibitem{drs}
\artref
{F.Delduc, E.Ragoucy and P.Sorba}{Super-\-Toda theo\-ries and 
$W$-\-alge\-bras from superspace Wess-\-Zumino-\-Witten 
models}{Commun.Math.Phys.}{146}{(1992) 403 .}

\bibitem{rs}
\bookref
{A.V.Razumov and M.V.Saveliev}{Lie algebras, Geometry and 
Toda-type Systems}{Cambridge Lecture Notes in Physics Vol.8
(Cambridge University Press, 1997) .}

\bibitem{japan}
\artref
{S. Komata, K. Mohri and H. Nohara}{Classical and quantum 
extended superconformal algebra}{Nucl.Phys.}{B359}{(1991) 
168-200 .}


\end{thebibliography}
\end{document}